\documentclass[12pt]{amsart}
\usepackage{amsmath}
\usepackage{amssymb}
\usepackage{amsthm}
\usepackage{epsfig}
\usepackage{euscript}
\usepackage{latexsym}
\usepackage{amsfonts}

\input{epsf}

\begin{document}

\newtheorem{fig}{Figure}[section]
\newtheorem{thm}{Theorem}[section]
\newtheorem{prp}[thm]{Proposition}
\newtheorem{lem}[thm]{Lemma}
\newtheorem{dfn}[thm]{Definition}
\newtheorem{cor}[thm]{Corollary}
\newtheorem{eg}[thm]{Example}
\newtheorem{rk}[thm]{Remark}
\parindent=0mm
\newcommand{\ndnt}{\hspace{7mm}}
\newcommand{\hs}{\hspace{2mm}}
\newcommand{\vs}{\vspace{1cm}}
\newcommand{\s}{\vspace{3mm}}
\newcommand{\as}{\vspace{5mm}}

\title{Fractal Strings and Multifractal Zeta Functions}
\author{Michel~L. Lapidus, Jacques L\'{e}vy-V\'{e}hel\\ and John~A.
Rock}

\date{\today}
\thanks{The work of MLL was partially supported by the US National
Science Foundation under contract DMS--070497}

\begin{abstract}
\noindent For a Borel measure on the unit interval and a sequence of
scales that tend to zero, we define a one-parameter family of zeta
functions called {\it multifractal zeta functions}. These functions
are a first attempt to associate a zeta function to certain
multifractal measures. However, we primarily show that they
associate a new zeta function, the {\it topological zeta function},
to a fractal string in order to take into account the topology of
its fractal boundary. This expands upon the geometric information
garnered by the traditional geometric zeta function of a fractal
string in the theory of complex dimensions. In particular, one can
distinguish between a fractal string whose boundary is the classical
Cantor set, and one whose boundary has a single limit point but has
the same sequence of lengths as the complement of the Cantor set.
Later work will address related, but somewhat different, approaches
to multifractals themselves, via zeta functions, partly motivated by
the present paper.
\end{abstract}

\maketitle

\pagebreak

\setcounter{section}{-1}
\section{Introduction}\label{intro}

Natural phenomena such as the distribution of ground water, the
formation of lightning and snowflakes, and the dissipation of
kinetic energy in turbulence can be modeled by multifractal
measures. Such measures can be described as a mass distribution
whose concentrations of mass vary widely when spread out over their
given regions. As a result, the region may be separated into
disjoint sets which are described by their Hausdorff dimensions and
are defined by their behavior with respect to the distribution of
mass. This distribution yields the {\it multifractal spectrum}: a
function whose output values are Hausdorff dimensions of the sets
which correspond to the input values. The multifractal spectrum is a
well-known tool in multifractal analysis and is one of the key
motivations for the {\it multifractal zeta functions} defined in
Section \ref{mzf}. Multifractal zeta functions were initially
designed to create another kind of multifractal spectrum which could
(potentially) be used to more precisely describe the properties of
multifractal measures. Although this paper does not accomplish this
feat, the main result is the gain of topological information for a
fractal subset of the real line: information which cannot be
obtained through use of the traditional geometric zeta function of
the corresponding fractal string (open complement of said fractal
subset) via the special case of multifractal zeta functions called
{\it topological zeta functions}. Further, multifractal zeta
functions provide the motivation for the zeta functions which appear
in \cite{LR,LR2,LVM,Rock} and are discussed in the epilogue of this
paper, Section \ref{epilogue}. Other approaches to multifractal
analysis can be found in
\cite{AP,BMP,CM,EM,Ellis,Falc,Ja1,Ja2,Ja3,Jaf,JaMey,LapvF5,LR,LR2,Lau,
LEV98a,LVM,LVR,LVS,LVV,BM,Man,Ol,O1,O2,PF,Rock}.

\ndnt For a measure and a sequence of scales, we define a family of
multifractal zeta functions parameterized by the extended real
numbers and investigate their properties. We restrict our view
mostly to results on fractal strings, which are bounded open subsets
of the real line. For a given fractal string, we define a measure
whose support is contained in the boundary of the fractal string.
This allows for the use of the multifractal zeta functions in the
investigation of the geometric and topological properties of fractal
strings. The current theory of geometric zeta functions of fractal
strings (see \cite{LapvF1,LapvF4}) provides a wealth of information
about the geometry and spectrum of these strings, but the
information is independent of the topological configuration of the
open intervals that comprise the strings. Under very mild
conditions, we show that the parameter $\alpha = \infty$ yields the
multifractal zeta function which precisely recovers the geometric
zeta function of the fractal string. Other parameter values are
investigated and, in particular, for certain measures and under
further conditions, the parameter $\alpha =-\infty$ yields a
multifractal zeta function, called the {\it topological zeta
function}, whose properties depend heavily on the topological
configuration of the fractal string in question.

\ndnt This paper is organized as follows:

\ndnt Section \ref{fsgzf} provides a brief review of fractal strings
and geometric zeta functions, along with a description of a few
examples which will be used throughout the paper, including the
Cantor String. Work on fractal strings can be found in
\cite{BesTa,HL,HeLap,Lap1,Lap2,Lap3,LapMa,LapPo1,LapPo2} and work on
geometric zeta functions and complex dimensions can be found in
\cite{LapvF1,LapvF2,LapvF3,LapvF4}.

\ndnt Section \ref{ma} provides a brief review of a tool and a
relatively simple example from multifractal analysis. The tool, {\it
regularity}, is integral to this paper, but the example, the
binomial measure, merely provides motivation and is not considered
again until the epilogue, Section \ref{epilogue}.

\ndnt Section \ref{mzf} contains the (lengthy) development and
definition of the main object of study, the multifractal zeta
function.

\ndnt Section \ref{rvigzf} contains a theorem describing the
recovery of the geometric zeta function of a fractal string for
parameter value $\alpha = \infty$.

\ndnt Section \ref{rvnitzf} contains a theorem describing the
topological configuration of a fractal string for parameter value
$\alpha = -\infty$ and the definition of topological zeta function.

\ndnt Section \ref{cs} investigates the properties of various
multifractal zeta functions for the Cantor String and a collection
of fractal strings which are closely related to the Cantor String.

\ndnt Section \ref{conclusion} concludes with a summary of the
results of this paper and points the interested reader in the
direction of other related topics such as higher-dimensional
fractals and multifractals as well as random fractal strings.

\ndnt Section \ref{epilogue} is an epilogue which discusses a few of
the results from \cite{LR,LR2,LVM,Rock}, concerning suitable
modifications of the multifractal zeta functions introduced in this
paper, specifically with regard to the binomial measure and its
multifractal spectrum, both of which are discussed briefly in
Section \ref{ma} below.

\section{Fractal Strings and Geometric Zeta Functions}\label{fsgzf}

In this section we review the current results on fractal strings,
geometric zeta functions and complex dimensions (all of which we
define below). Results on fractal strings can be found in
\cite{BesTa,HL,HeLap,Lap1,Lap2,Lap3,LapMa,LapPo1,LapPo2} and results
on geometric zeta functions and complex dimensions can be found in
\cite{LapvF1,LapvF2,LapvF3,LapvF4}.

\begin{dfn}\label{def:fs}
A \underline{fractal string} $\Omega$ is a bounded open subset of
the real line.
\end{dfn}

\begin{figure}
\epsfysize=6.5cm\epsfbox{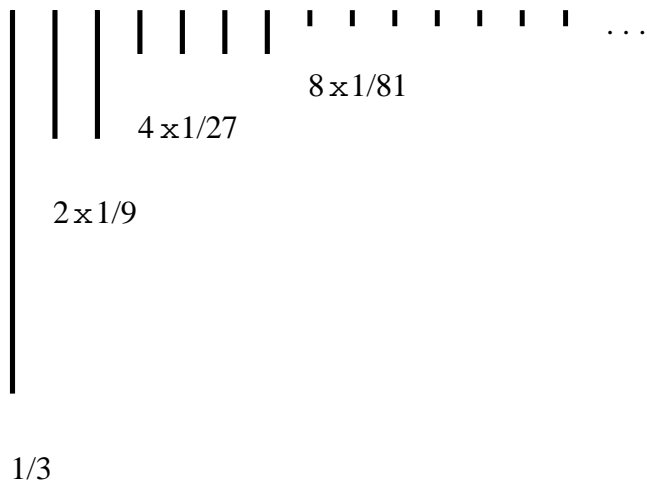}
    \caption{\textit{The lengths of the Cantor String.}}
\end{figure}

Unlike \cite{LapvF1,LapvF4}, it will be necessary to distinguish
between a fractal string $\Omega$ and its sequence of lengths
$\mathcal{L}$ (with multiplicities). That is, the sequence \(
\mathcal{L}=\{\ell_j\}_{j=1}^{\infty}\) is the nonincreasing
sequence of lengths of the disjoint open intervals \( I_j =
(a_j,b_j)\) where \(\Omega = \cup_{j=1}^{\infty} I_j.\) (Hence, the
intervals $I_j$ are the connected components of $\Omega$.)We will
need to consider the sequence of \textit{distinct} lengths, denoted
$\{l_n\}_{n=1}^{\infty}$, and their multiplicities
$\{m_n\}_{n=1}^{\infty}$. Two useful examples of fractal strings are
the $a$-String and the Cantor String, both of which can be found in
\cite{LapvF1,LapvF4}. The lengths of the Cantor String appear in
Figure 1.

\ndnt Below we recall a generalization of Minkowski dimension
called \textit{complex dimensions} which are used to study the
properties of certain fractal subsets of $\mathbb{R}$. For instance,
the boundary of a fractal string $\Omega$, denoted $\partial\Omega$,
can be studied using complex dimensions.

\ndnt Let us now describe some preliminary notions. We take $\Omega$
to be a fractal string and $\mathcal{L}$ its associated sequence of
lengths. The one-sided volume of the tubular neighborhood of radius
$\varepsilon$ of $\partial\Omega$ is
\[
V(\varepsilon)=\lambda(\{x \in \Omega \hs | \hs
dist(x,\partial\Omega)<\varepsilon\}),
\]

where $\lambda(\cdot)=|\cdot|$ denotes the Lebesgue measure. The
\textit{Minkowski dimension} of $\mathcal{L}$ is
\[
D=D_{\mathcal{L}}:=\inf \{\alpha \geq 0 \hs | \hs
\limsup_{\varepsilon \rightarrow
0^{+}}V(\varepsilon)\varepsilon^{\alpha-1} <\infty \}.
\]

Note that we refer directly to the sequence $\mathcal{L}$, not the
boundary of $\Omega$, due to the translation invariance of the
Minkowski dimension.

\ndnt If \(\lim_{\varepsilon \rightarrow
0^{+}}V(\varepsilon)\varepsilon^{\alpha-1}\) exists and is positive
and finite for some $\alpha$, then $\alpha = D$ and we say that
$\mathcal{L}$ is \textit{Minkowski measurable}. The
\textit{Minkowski content} of $\mathcal{L}$ is then defined by
\(\mathcal{M}(D,\mathcal{L}):=\lim_{\varepsilon \rightarrow
0^{+}}V(\varepsilon)\varepsilon^{D-1}.\)

\ndnt The Minkowski dimension is also known as the
\textit{box-counting dimension} because, for a bounded subset $F$ of
$\mathbb{R}^d$, it can also be expressed in terms of
\[
\limsup_{\varepsilon \rightarrow
0^{+}}\frac{N_{\varepsilon}(F)}{-\log{\varepsilon}},
\]
where $N_{\varepsilon}(F)$ is the smallest number of cubes with side
length $\varepsilon$ that cover $F$. In \cite{Lap1}, it is shown
that if $F = \partial\Omega$ is the boundary of a bounded open set
$\Omega$, then \(d-1 \leq dim_H(F) \leq D \leq d\) where $d$ is the
dimension of the ambient space, $dim_H(F)$ is the Hausdorff
dimension of $F$ and $D=dim_{M}(F)$ is the Minkowski dimension of
$F$ (with ``1'' replaced by ``d'' in the above definition). In
particular, in this paper, we have $d=1$ and hence \(0 \leq dim_H(F)
\leq D \leq 1.\)

\ndnt The following equality describes an interesting relationship
between the Minkowski dimension of a fractal string $\Omega$ (really
the Minkowski dimension of $\partial\Omega$) and the sum of each of
its lengths with exponent $\sigma \in \mathbb{R}$. This was first
observed in \cite{Lap2} using a key result of Besicovitch and Taylor
\cite{BesTa}, and a direct proof can be found in \cite{LapvF4}, pp.
17--18:

\[
D=D_{\mathcal{L}}=\inf \bigg\{ \sigma \geq 0 \hs | \hs
\sum_{j=1}^{\infty}\ell_{j}^{\sigma} <\infty \bigg\}.
\]

We can consider $D_{\mathcal{L}}$ to be the abscissa of convergence
of the Dirichlet series \(\sum_{j=1}^{\infty}\ell_{j}^{s}\), where
$s \in \mathbb{C}$. This Dirichlet series is the \textit{geometric
zeta function} of $\mathcal{L}$ and it is the function that we will
generalize using notions from multifractal analysis.

\begin{dfn}\label{def:gzf} The \underline{geometric zeta function}
of a fractal string $\Omega$ with lengths $\mathcal{L}$ is
\[
\zeta_{\mathcal{L}}(s)=\sum_{j=1}^{\infty}\ell_{j}^{s}=
\sum_{n=1}^{\infty}m_{n}l_{n}^{s},
\] where
$\textnormal{Re}(s)>D_{\mathcal{L}}$.
\end{dfn}
We may consider lengths $\ell_j=0$, in which case we use the
convention that $0^s =0$ for all $s \in \mathbb{C}$.

\ndnt One can extend the notion of the dimension of a fractal string
$\Omega$ to complex values by considering the poles of
$\zeta_{\mathcal{L}}$. In general, $\zeta_{\mathcal{L}}$ may not
have an analytic continuation to all of $\mathbb{C}$. So we consider
regions where $\zeta_{\mathcal{L}}$ has a meromorphic extension and
collect the poles in these regions. Specifically, consider the
\textit{screen} $S$ where
\[
S=r(t)+it,
\] for some continuous
function \( r: \mathbb{R} \rightarrow [-\infty, D_{\mathcal{L}}]\)
and consider the \textit{window} $W$ which are the complex numbers
to the right of the screen. That is,
\[
W=\{s \in \mathbb{C} \hs | \hs \textnormal{Re}(s) \geq
r(\textnormal{Im}(s)) \}.
\]
Assume that $\zeta_{\mathcal{L}}$ has a meromorphic extension to an
open neighborhood of $W$ and there is no pole of
$\zeta_{\mathcal{L}}$ on $S$.

\begin{dfn}\label{def:cd}
The set of \underline{complex dimensions} of a fractal string
$\Omega$ with lengths $\mathcal{L}$ is
\[
\mathcal{D}_{\mathcal{L}}(W)=\{\omega \in W \hs | \hs
\zeta_{\mathcal{L}} \hs \textnormal{has a pole at } \omega\}.
\]
\end{dfn}

The following is a result characterizing Minkowski measurability
which can be found in \cite{LapvF1,LapvF4}.

\begin{thm}\label{thm:cdmink}
If a fractal string $\Omega$ with lengths $\mathcal{L}$ satisfies
certain mild conditions, the following are equivalent:
\begin{enumerate}
\item $D$ is the only complex dimension of $\Omega$ with real part
$D_{\mathcal{L}}$, and it is simple.
\item $\partial \Omega$ is Minkowski measurable.
\end{enumerate}
\end{thm}

The above theorem applies to all self-similar strings, including the
Cantor String discussed below.

\ndnt Earlier, the following criterion was obtained in
\cite{LapPo1}.

\begin{thm}\label{thm:Lmink}
Let $\Omega$ be an arbitrary fractal string with lengths
$\mathcal{L}$ and $0 < D< 1$. The following are equivalent:
\begin{enumerate}
\item \(L:=\lim_{j \rightarrow \infty} \ell_j \cdot j^{1/D}\) exists
in \( (0,\infty). \)
\item $\partial\Omega$ is Minkowski measurable.
\end{enumerate}
\end{thm}

\begin{rk}\label{rk:either} \textnormal{When one of the conditions of either theorem is satisfied, the
Minkowski content of $\mathcal{L}$ is given by}
\end{rk}
\[
\mathcal{M}(D,\mathcal{L})= \frac{2^{1-D}L^D}{1-D}.
\]
Further, under the conditions of Theorem \ref{thm:cdmink}, we also
have
\[
\mathcal{M}(D,\mathcal{L})= \textnormal{res}(\zeta_{\mathcal{L}};D).
\]

\begin{eg}[\textbf{Cantor String}]\label{eg:cs}
\textnormal{Let $\Omega_1$ be the Cantor String, defined as the
complement in $[0,1]$ of the ternary Cantor Set, so that
$\partial\Omega_1$ is the Cantor Set itself. (See Figure 2.) The
distinct lengths are $l_n = 3^{-n}$ with multiplicities $m_n =
2^{n-1}$ for every $n \geq 1$. Hence,
\begin{eqnarray*}
\zeta_{\mathcal{L}}(s) &=& \sum_{n=1}^{\infty}m_nl_n^s
=\sum_{n=1}^{\infty}2^{n-1}3^{-ns}\\
&=& \frac{3^{-s}}{1-2 \cdot 3^{-s}}, \hs \textnormal{for Re}(s) >
\frac{ \log2}{\log3}.
\end{eqnarray*}
Upon meromorphic continuation, we see that
\[
\zeta_{\mathcal{L}}(s) = \frac{3^{-s}}{1-2 \cdot 3^{-s}}, \hs
\textnormal{for all } s \in \mathbb{C},
\]
and hence
\[
\mathcal{D}_{\mathcal{L}} = \left\{ \log_2{3} + \frac{2im\pi}{\log3}
\hs | \hs m \in \mathbb{Z} \right\}.
\]
Note that $D_{\mathcal{L}} = \log_2{3}$ is the Minkowski dimension,
as well as the Hausdorff dimension, of the Cantor Set
$\partial\Omega_1$. Thus by Theorem \ref{thm:cdmink}, the Cantor Set
is not Minkowski measurable. The latter fact can also be deduced
from Theorem \ref{thm:Lmink}, as was first shown in \cite{LapPo1}.}
\end{eg}

\begin{figure}
\epsfysize=2.2cm\epsfbox{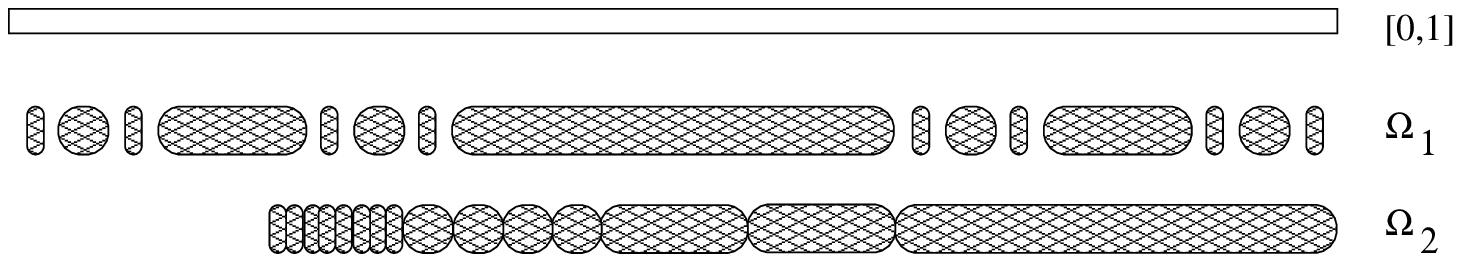} \caption{\textit{The
first four distinct lengths, with multiplicities, of the Cantor
String} $\Omega_1$ \textit{and the fractal string} $\Omega_2$.}
\end{figure}

\begin{eg}[\textbf{A String with the Lengths of the Cantor
String}]\label{eg:cs2} \textnormal{Let $\Omega_2$ be the fractal
string that has the the same lengths as the Cantor String, but with
the lengths arranged in non-increasing order from right to left.
(See Figure 2.) This fractal string has the same geometric zeta
function as the Cantor String, and thus the same Minkowski
dimension, $\log_{3}2$; however, the Hausdorff dimension of the
boundary of $\Omega_2$ is zero, whereas that of $\Omega_1$ is
$\log_{3}2$ (by the self-similarity of the Cantor Set, see
\cite{Falc}). This follows immediately from the fact that the
boundary is a set of countably many points. The multifractal zeta
functions defined in Section \ref{mzf} below will illustrate this
difference and hence allow us to distinguish between the fractal
strings $\Omega_1$ and $\Omega_2$.}
\end{eg}

\ndnt The following key result, which can be found in
\cite{LapvF1,LapvF4}, uses the complex dimensions of a fractal
string in a formula for the volume of the inner
$\varepsilon$-neighborhoods of a fractal string.

\begin{thm}\label{thm:vol}
Under mild hypotheses, the volume of the one-sided tubular
neighborhood of radius $\varepsilon$ of the boundary of a fractal
string $\Omega$ (with lengths $\mathcal{L}$) is given by the
following explicit formula with error term:

\[
V(\varepsilon) = \sum_{\omega \in \mathcal{D}_{\mathcal{L}}(W)\cup
\{0\}} \textnormal{res}\left(\frac{\zeta_{\mathcal{L}}
(s)(2\varepsilon)^{1-s}}{s(1-s)};\omega \right) +
\mathcal{R}(\varepsilon),
\]
where the error term can be estimated by \( \mathcal{R}(\varepsilon)
= \mathcal{O}(\varepsilon^{1-\sup r}) \) as \( \varepsilon
\rightarrow 0^+.\)

\end{thm}

\begin{rk} \textnormal{In particular, in Theorem \ref{thm:vol},
if all the poles of} $\zeta_{\mathcal{L}}$ \textnormal{are simple
and} \( 0 \notin \mathcal{D}_{\mathcal{L}}(W),\) \textnormal{then}
\end{rk}
\[
V(\varepsilon) = \sum_{\omega \in \mathcal{D}_{\mathcal{L}}(W)}
\frac{2^{1-\omega}}{\omega(1-\omega)}\textnormal{res}(\zeta_{\mathcal{L}},
\omega)\varepsilon^{1-\omega} + \mathcal{R}(\varepsilon).
\]

\begin{rk} \textnormal{If $\mathcal{L}$ is a self-similar string}
\textnormal{(e.g., if its boundary is a self-similar subset of
$\mathbb{R}$), then the conclusion of Theorem \ref{thm:vol} holds
with} \( \mathcal{R}(\varepsilon) \equiv 0.\) \textnormal{This is
the case, in particular, for the Cantor String $\Omega_1$ and for
$\Omega_2$ discussed in Examples \ref{eg:cs} and \ref{eg:cs2}.}
\end{rk}

\section{Multifractal Analysis}\label{ma}

Multifractal analysis is the study of measures which can be
described as mass distributions whose concentrations of mass vary
widely. In this section and throughout this text, we restrict our
view to measures on the unit interval $[0,1]$. Example \ref{eg:mmcs}
below briefly discusses the construction of a multifractal measure
on the Cantor set, with Figure 3 providing a few steps of the
construction and the resulting multifractal spectrum, as found in
Chapter 17 of \cite{Falc}.

\begin{figure}
\epsfysize=4cm\epsfbox{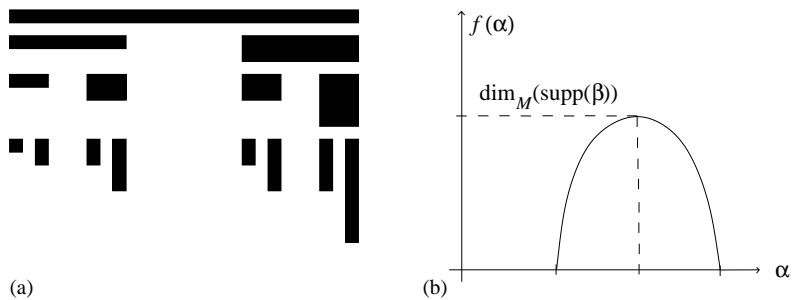}
    \caption{\textit{\textnormal{(a)} Construction of the binomial
    measure $\beta$. \textnormal{(b)} The multifractal spectrum
    $f(\alpha)$ of the measure $\beta$.}}
\end{figure}

\begin{eg}[\textbf{A multifractal measure on the Cantor
set}]\label{eg:mmcs} \textnormal{A simple example of a multifractal
measure is the {\it binomial measure} $\beta$ constructed on the
classical Cantor set. To construct $\beta$, a mass distribution is
added to the construction of the Cantor set which consists of a
countable intersection of a nonincreasing sequence of closed
intervals whose lengths tend to zero. Specifically, in addition to
removing open middle thirds, weight is added at each stage. On the
remaining closed intervals of each stage of the construction, place
$1/3$ of the weight on the left interval and $2/3$ on the right, ad
infinitum (see Figure 3). The measure found in the limit, denoted
$\beta$, is a multifractal measure.}
\end{eg}

\ndnt A notion which is key to the development of the multifractal
zeta functions and is part of Example \ref{eg:mmcs} is {\it
regularity}. Regularity connects the size of a set with its mass.
Specifically, let \( \textbf{X}([0,1]) \) denote the space of closed
subintervals of \([0,1]\). The following definition can be found in
\cite{LVT} and is motivated by the large deviation spectrum and one
of the continuous large deviation spectra in that paper.

\begin{dfn}\label{def:reg}
The \underline{regularity} \(A(U)\) of a Borel measure $\mu$ on \( U
\in \textnormal{\textbf{X}([0,1])} \) is
\[
A(U)=\frac{\log \mu(U)}{\log{|U|}},
\]
where \( |\cdot| = \lambda(\cdot) \) is the Lebesgue measure on
$\mathbb{R}$.
\end{dfn}
Regularity $A(U)$ is also known as the {\it coarse H\"older
exponent}  $\alpha$ which satisfies
\[
|U|^{\alpha}=\mu(U).
\]
We will consider regularity values $\alpha$ in the extended real
numbers $[-\infty,\infty]$, where

\[
\alpha=\infty=A(U) \Leftrightarrow \mu(U) =0 \textnormal{ and } |U|
> 0,
\]
and
\[
\alpha=-\infty=A(U) \Leftrightarrow \mu(U) = \infty  \textnormal{
and } |U| > 0.
\]

\ndnt The original motivation for using the regularity values was to
develop a family of zeta functions which would be parameterized by
these values and would, in turn, generate a family of complex
dimensions such as those from Section \ref{fsgzf}, but now indexed
by the regularity exponent $\alpha \in [-\infty, \infty]$. A new
kind of multifractal spectrum was to then be developed where the
function values $f(\alpha)$ would be the real-valued dimensions (of
some sort) corresponding to the regularity values $\alpha$. However,
the multifractal zeta functions defined and discussed in the
following sections are quite complicated and their application to
examples such as Example \ref{eg:mmcs} has yet to be thoroughly
examined. Our view is restricted to measures that can be
characterized as a collection of point-masses on the boundary of a
fractal string. The results, therefore, are not rich enough to
generate a full spectrum of dimension for the measures we consider.
Nevertheless, multifractal zeta functions do regenerate the
classical geometric zeta functions and reveal new topological zeta
functions for fractal strings when examined via our restricted
collection of measures. Analysis of multinomial measures using other
families of zeta functions, whose definitions were motivated in part
by those defined in the next section, has been done in
\cite{LR,LR2,LVM,Rock}. For other approaches to multifractal
analysis, consider
\cite{AP,BMP,CM,EM,Ellis,Falc,Ja1,Ja2,Ja3,Jaf,JaMey,LapvF5,LR,LR2,Lau,LEV98a,LVR,LVS,LVV,BM,Man,Ol,O1,O2,PF}.

\section{Multifractal Zeta Functions}\label{mzf}

In order to define multifractal zeta functions, we must understand
the behavior of the measures with respect to their regularity values
in significant detail. As such, more tools are provided below before
the definition is given.

\ndnt Let the collection of closed intervals with length $\eta \in
(0,1)$ and regularity $\alpha$ be denoted by
$\mathcal{R}_{\eta}(\alpha)$. Namely,
\[
\mathcal{R}_{\eta}(\alpha) = \{U \in \textbf{X}([0,1]) \hs | \hs
|U|=\eta \textnormal{ and } A(U)=\alpha\}.
\]

Consider the union of the sets in $\mathcal{R}_{\eta}(\alpha)$,
\[
\bigcup_{ \mathcal{R}_{\eta}(\alpha)}U := \bigcup_{U \in
\mathcal{R}_{\eta}(\alpha)}U.
\]

Given $\alpha \in [-\infty,\infty]$ and $\eta \in (0,1)$, let
\[
 R^{\eta}(\alpha) = \bigcup_{ \mathcal{R}_{\eta}(\alpha)}U.
\]
For a scale \( \eta> 0\), $R^{\eta}(\alpha)$ is a disjoint union of
a finite number of intervals, each of which may be open, closed or
neither and are of length at least $\eta$ when
$\mathcal{R}_{\eta}(\alpha)$ is non-empty. We will consider only
discrete sequences of scales \( \mathcal{N} =
\{\eta_n\}_{n=1}^{\infty}\), with $\eta_n > 0$ for all $n \geq 1$
and the sequence strictly decreasing to zero. So for $n \in
\mathbb{N}$, let
\[
R^{\eta_n}(\alpha)=R^{n}(\alpha).
\]

We have

\[R^{n}(\alpha)=\bigcup_{p=1}^{r_{n}(\alpha)}R^{n}_{p}(\alpha),\]

where $r_{n}(\alpha)$ is the number of connected components
$R^{n}_{p}(\alpha)$ of $R^{n}(\alpha)$. We denote the left and right
endpoints of each interval $R^{n}_{p}(\alpha)$ by
\(a_R^n(\alpha,p)\) and \(b_R^n(\alpha,p),\) respectively.

\ndnt Given a sequence of positive real numbers \( \mathcal{N}=
\{\eta_{n}\}_{n=1}^{\infty} \) that tend to zero and a Borel measure
$\mu$ on [0,1], we wish to examine the way $\mu$ changes with
respect to a fixed regularity $\alpha$ between stages $n-1$ and $n$.
Thus we consider the symmetric difference ($\ominus$) between
$R^{n-1}(\alpha)$ and $R^{n}(\alpha)$. Let $J^{1}(\alpha) =
R^1(\alpha),$ and for $n \geq 2$, let
\[
J^{n}(\alpha) = R^{n-1}(\alpha) \ominus R^{n}(\alpha).
\]

For all $n\in \mathbb{N}$, $J^{n}(\alpha)$ is also a disjoint union
of intervals $J^{n}_{p}(\alpha)$, each of which may be open, closed,
or neither. We have
\[J^{n}(\alpha)= \bigcup_{p=1}^{j_{n}(\alpha)}J^{n}_{p}(\alpha), \]

where $j_{n}(\alpha)$ is the number of connected components
$J^{n}_{p}(\alpha)$ of $J^{n}(\alpha)$. The left and right endpoints
of each interval $J^{n}_{p}(\alpha)$ are denoted by
\(a_J^n(\alpha,p)\) and \(b_J^n(\alpha,p),\) respectively.

\ndnt For a given regularity \(\alpha \in [-\infty,\infty] \) and a
measure $\mu$, the sequence $\mathcal{N}$ determines another
sequence of lengths corresponding to the lengths of the connected
components of the $J^{n}(\alpha)$. That is, the $J^{n}(\alpha)$
describe the way $\mu$ behaves between scales $\eta_{n-1}$ and
$\eta_{n}$ with respect to $\alpha$. However, there is some
redundancy with this set-up. Indeed, a particular regularity value
may occur at all scales below a certain fixed scale in the same
location. The desire to eliminate this redundancy will be clarified
with some examples below. The next step is introduced to carry out
this elimination.

\ndnt Let $K^1(\alpha) = J^{1}(\alpha) = R^1(\alpha)$. For $n \geq
2$, let $K^n(\alpha)$ be the union of the subcollection of intervals
in $J^n(\alpha)$ comprised of the intervals that have left and right
endpoints distinct from, respectively, the left and right endpoints
of the intervals in $R^{n-1}(\alpha)$. We have

\[
K^{n}(\alpha)= \bigcup_{p=1}^{k_{n}(\alpha)}K^{n}_{p}(\alpha)
\subset J^n(\alpha),
\]

where $k_{n}(\alpha)$ is the number of connected components
$K^{n}_{p}(\alpha)$ of $K^{n}(\alpha)$. That is, the
$K^{n}_{p}(\alpha)$ are the $J^{n}_p(\alpha)$ such that
\(a^n_J(\alpha,p_1) \neq a^{n-1}_R(\alpha,p_2) \) and
\(b^n_J(\alpha,p_1) \neq b^{n-1}_R(\alpha,p_2) \) for all \( p_1 \in
\{1,...,j_n(\alpha)\}\) and \( p_2 \in \{1,...,r_n(\alpha)\}.\)
Collecting the lengths of the intervals $K^n_p(\alpha)$ allows one
to define a generalization of the geometric zeta function of a
fractal string by considering a family of geometric zeta functions
parameterized by the regularity values of the measure $\mu$.

\begin{dfn}\label{def:mzf} The \underline{multifractal zeta function}
of a measure $\mu$, sequence $\mathcal{N}$ and \underline{with
associated regularity value $\alpha$} $\in [-\infty,\infty]$ is
\[
\zeta^{\mu}_{\mathcal{N}}(\alpha,s) =
\sum_{n=1}^{\infty}\sum_{p=1}^{k_{n}(\alpha)}|K^{n}_{p}(\alpha)|^s,
\]
for \textnormal{Re}$(s)$ large enough.
\end{dfn}

\ndnt If we assume that, as a function of $s \in \mathbb{C}$,
\(\zeta^{\mu}_{\mathcal{N}}(\alpha,s) \) admits a meromorphic
continuation to an open neighborhood of a window $W$, then we may
also consider the poles of these zeta functions, as in the case of
the complex dimensions of a fractal string (see Section
\ref{fsgzf}).

\begin{dfn}\label{def:poles} For a measure $\mu$, sequence
$\mathcal{N}$ which tends to zero and regularity value $\alpha$, the
\underline{set of complex dimensions with parameter $\alpha$} is
given by
\[
\mathcal{D}^{\mu}_{\mathcal{N}}(\alpha, W) = \{\omega \in W \hs |
\hs \zeta^{\mu}_{\mathcal{N}}(\alpha,s) \textnormal{ has a pole at }
\omega \}.
\]
When $W =\mathbb{C}$, we simply write
$\mathcal{D}^{\mu}_{\mathcal{N}}(\alpha)$.
\end{dfn}

\ndnt The following sections consider two specific regularity
values. In Section \ref{rvigzf}, the value $\infty$ generates the
geometric zeta function for the complement of the support of the
measure in question. In Section \ref{rvnitzf}, the value $-\infty$
generates the topological zeta function which detects some
topological properties of fractal strings that are ignored by the
geometric zeta functions when certain measures are considered.

\section{Regularity Value $\infty$ and Geometric Zeta
Functions}\label{rvigzf}

The geometric zeta function is recovered as a special case of
multifractal zeta functions. Specifically, regularity value
$\alpha=\infty$ yields the geometric zeta function of the complement
of the support of a given positive Borel measure $\mu$ on $[0,1]$.

\ndnt To see how this is done, let $E^c$ denote the complement of
$E$ in $[0,1]$ and consider the fractal string $(supp(\mu))^{c} =
\Omega_{\mu}$ whose lengths $\mathcal{L}_{\mu}$ are those of the
disjoint intervals \((a_{j},b_{j}) \) where \(\Omega_{\mu}=
\cup_{j=1}^{\infty}(a_{j},b_{j})\). Let
$\{\ell_{j}\}_{j=1}^{\infty}$ be the lengths of $\mathcal{L}_{\mu}$.
Thus,
\[
\{\ell_{j}\}_{j=1}^{\infty} =|(a_{j},b_{j})| = b_j - a_j.
\]

Further, let $\{l_n\}_{n=1}^{\infty}$ be the distinct lengths of
$\mathcal{L}_{\mu}$ with multiplicities $\{m_{n}\}_{n=1}^{\infty}$.

\begin{figure}
\epsfysize=4.5cm\epsfbox{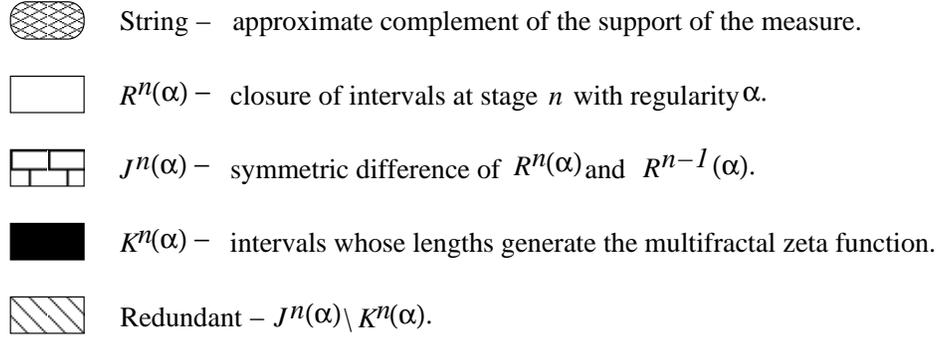}
    \caption{\textit{Key for the construction of the lengths used to define
    the multifractal zeta functions.}}
\end{figure}

\ndnt The following technical lemma is used in the proof of the
theorem below which shows the recovery of the geometric zeta
function as the multifractal zeta function with regularity $\infty$.
See Figures 4 and 5 for an illustration of the construction of a
multifractal zeta function with regularity $\infty$ for a measure
which is supported on the Cantor set.

\begin{lem}\label{lemma}
Suppose \(\{x\}=supp(\mu) \cap U\) for some \(U \in
\textnormal{\textbf{X}}([0,1])\). Then
\[A(U)= \infty \Leftrightarrow \mu(\{x\})=0.\]
\end{lem}

\begin{proof} \( \mu(\{x\}) \neq 0 \Leftrightarrow \mu(U) > |U|
\Leftrightarrow A(U) \neq \infty. \)
\end{proof}

\ndnt The lemma helps deal with the subtle interactions between the
closed intervals $U$ of size $\eta_n$ and the support of $\mu$,
essentially allowing us to prove a single case of the following
theorem without loss of generality.

\begin{figure}
\epsfysize=11cm\epsfbox{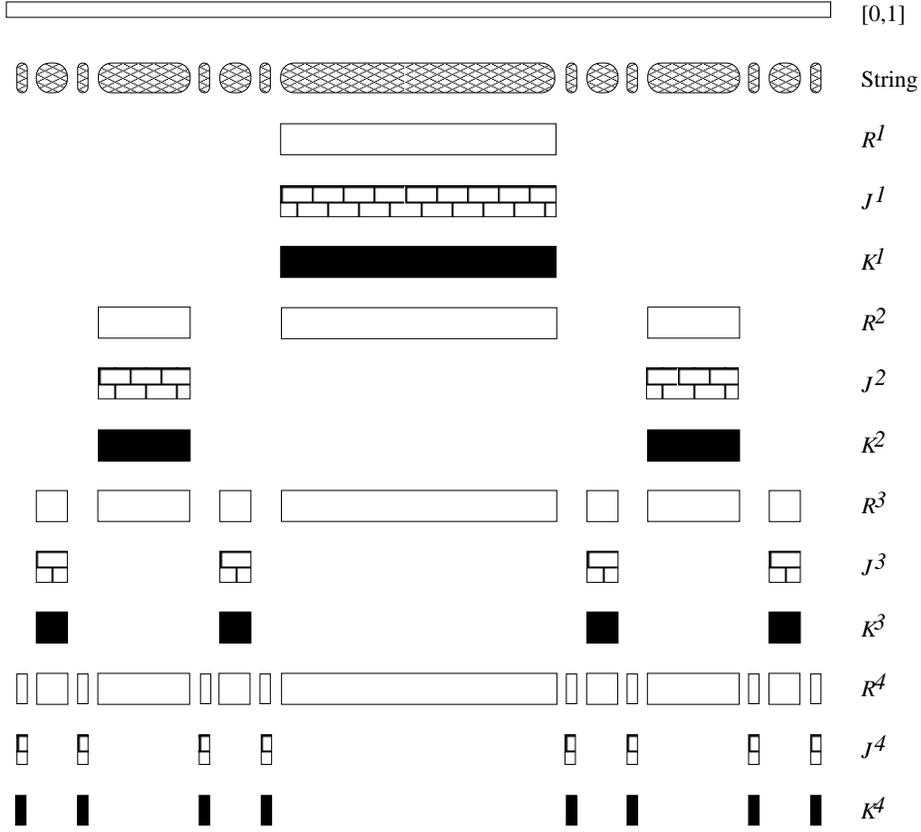}
    \caption{\textit{Construction of the
multifractal zeta function} $\zeta^{\mu}_{\mathcal{N}}(\infty,s)$
\textit{as in the proof of Theorem} \ref{thm:recovergzf}.}
\end{figure}

\begin{thm}\label{thm:recovergzf}
The multifractal zeta function of a positive Borel measure $\mu$,
any sequence $\mathcal{N}$ such that \(\eta_{n} \searrow 0\) and
regularity $\alpha = \infty$ is the geometric zeta function of
$(supp(\mu))^c$. That is,
\[
\zeta^{\mu}_{\mathcal{N}}(\infty,s) = \zeta_{\mathcal{L}_{\mu}}(s).
\]
\end{thm}

\begin{proof}
Recall the notation introduced at the beginning of Section
\ref{mzf}. For all \(n \in \mathbb{N}\),
\[
U \in \mathcal{R}_{\eta_{n}}(\infty) \Leftrightarrow A(U) =  \frac{\log(\mu(U))}{\log|U|} = \infty
\hs \textnormal{and} \hs |U|=\eta_{n}.
\]
Therefore, \( \forall n \in \mathbb{N}, U \in
\mathcal{R}_{\eta_{n}}(\infty)\) only if \(\mu(U)=0\).

\ndnt The sets $\mathcal{R}_{\eta_{n}}(\infty)$ depend further upon
whether any of the endpoints of the intervals $I_j = (a_j,b_j)$
which comprise $\Omega_{\mu} = (supp(\mu))^c$ contain mass as
singletons. If $\mu(\{a_j\}) \neq 0$ and $\mu(\{b_j\}) \neq 0$ for
all $j \in \mathbb{N}$, then \( R^n(\infty)=\bigcup_{\ell_j >
\eta_n}I_j \subset \Omega_{\mu}. \)

\ndnt Lemma \ref{lemma} implies that, without loss of
generality, we need only consider the case
where every endpoint contains mass. Suppose $\mu(\{a_j\}) \neq 0$ and $\mu(\{b_j\}) \neq 0$ for
all $j \in \mathbb{N}$. Then \( R^n(\infty)=\bigcup_{\ell_j >
\eta_n}I_j\) implies that, for $n \geq 2$,

\begin{eqnarray*}
J^n(\infty)&=& \Bigg(\bigcup_{\ell_j > \eta_{n-1}}I_j\Bigg) \ominus
    \Bigg(\bigcup_{\ell_j > \eta_n}I_j\Bigg)\\
&=& \Bigg(\bigcup_{\ell_j > \eta_{n}}I_j\Bigg) \setminus
    \Bigg(\bigcup_{\ell_j > \eta_{n-1}}I_j\Bigg)\\
&=& \bigcup_{\eta_{n-1} \geq \ell_j > \eta_{n}}I_j.
\end{eqnarray*}

Since $R^{n-1}(\infty) \subset R^n(\infty)$ for all $n \geq 2$, the
intervals $J^n(\infty)$ have no redundant lengths. That is,
$a_J^n(\infty,p_1) \neq a_R^{n-1}(\infty,p_2)$ and
$b_J^n(\infty,p_1) \neq b_R^{n-1}(\infty,p_2)$ for all $n \geq 2$
and $p_1, p_2 \in \{1,...,j_n(\infty)\}$. This implies
\[
K^n(\infty)=J^n(\infty)=\bigcup_{\eta_{n-1} \geq \ell_j >
\eta_{n}}I_j.
\]
Furthermore,
\[
|K^n(\infty)| = \sum_{p=1}^{k_n(\infty)}|K^n_p(\infty)| = \sum
\ell_j,
\]
where the last sum is taken over all $j$ such that $\eta_{n-1} \geq
\ell_j > \eta_{n}$. Since $\eta_n \searrow 0$, each length $\ell_j$
is eventually picked up. Therefore,

\begin{eqnarray*}
\zeta^{\mu}_{\mathcal{N}}(\infty,s) &=& \sum_{n=1}^{\infty}
\sum_{p=1}^{k_n(\infty)}|K^n_p(\infty)|^s
= \sum_{n=1}^{\infty} \sum \ell_j^s\\
&=& \sum_{n=1}^{\infty}m_nl_n^s
= \zeta_{\mathcal{L}_{\mu}}(s).
\end{eqnarray*}
\end{proof}

\begin{cor}
Under the assumptions of Theorem \ref{thm:recovergzf}, the complex
dimensions of the fractal string $\Omega_{\mu}=(supp(\mu))^c$
coincide with the poles of the multifractal zeta function
$\zeta^{\mu}_{\mathcal{N}}(\infty,s)$. That is,
\[
\mathcal{D}_{\mathcal{N}}^{\mu}(\infty,W)=
\mathcal{D}_{\mathcal{L_{\mu}}}(W)
\]
for every window $W$.
\end{cor}

\ndnt The key in Figure 4 will be used for the examples that analyze
the fractal strings below. Figure 5 shows the first four steps in
the construction of a multifractal zeta function with regularity
$\infty$ for a measure supported on the Cantor set.

\begin{rk} \textnormal{Assume $supp(\mu)$ has empty interior, as is
the case, for example, if $supp(\mu)$ is a Cantor set. It then
follows from Theorem \ref{thm:recovergzf} that }
$D_{\mathcal{L}_{\mu}}$, \textnormal{the abscissa of convergence of
} \(\zeta^{\mu}_{\mathcal{N}}(\infty,s)\), \textnormal{is the
Minkowski dimension of } \( \partial\Omega_{\mu} = supp(\mu). \)
\textnormal{ Note that as long as the sequence decreases to zero,
the choice of sequence of scales }$\mathcal{N}$ \textnormal{does not
affect the result of Theorem \ref{thm:recovergzf}. This is not the
case, however, for other regularity values.}
\end{rk}

\ndnt The following section describes a measure which is designed to
illuminate properties of a given fractal string and justifies
calling the multifractal zeta function with regularity $-\infty$ the
topological zeta function.

\section{Regularity Value $-\infty$ and Topological Zeta
Functions}\label{rvnitzf}

The remainder of this paper deals with fractal strings that have a
countably infinite number of lengths. If there are only a finite
number of lengths, it can be easily verified that all of the
corresponding zeta functions are entire because the measures taken
into consideration are then comprised of a finite number of unit
point-masses. Thus we consider certain measures that have infinitely
many unit point-masses. More specifically, in this section we
consider a fractal string $\Omega$ to be a subset of $[0,1]$
comprised of countably many open intervals $(a_j,b_j)$ such that
$|\Omega|=1$ and $\partial\Omega = [0,1]\backslash \Omega$ (or
equivalently, $\Omega^c = [0,1]\backslash \Omega$ has empty
interior). We also associate to $\Omega =
\cup_{j=1}^{\infty}(a_j,b_j)$ its sequence of lengths $\mathcal{L}$.
For such $\Omega$, the endpoints of the intervals $(a_j,b_j)$ are
dense in $\partial\Omega$. Indeed, if there were a point in
$\partial\Omega$ away from any endpoint, then it would be away from
$\Omega$ itself, meaning it would not be in $\partial\Omega$. This
allows us to define, in a natural way, measures with a countable
number of point-masses contained in the boundary of $\Omega$. Let
\[
\mu_{\Omega}:=\sum_{j=1}^{\infty}(\delta_{a_j}+\delta_{b_j}),
\]
where, as above, the $(a_j,b_j)$ are the open intervals whose
disjoint union is $\Omega$.

\ndnt Let us determine the nontrivial regularity values $\alpha$.
For $\alpha = \infty$, $\mathcal{R}_{\eta_n}(\infty)$ is the
collection of closed intervals of length $\eta_n$ which contain no
point-masses. For $\alpha = -\infty$,
$\mathcal{R}_{\eta_n}(-\infty)$ is the collection of closed
intervals of length $\eta_n$ which contain infinitely many
point-masses. In other words, $\mathcal{R}_{\eta_n}(-\infty)$ is the
collection of closed intervals of length $\eta_n$ that contain a
neighborhood of an accumulation point of the endpoints of $\Omega$.
This connection motivates the following definition.

\begin{dfn}\label{def:tzf} Let $\Omega$ be a fractal string and
consider the corresponding measure $\mu_{\Omega} =
\sum_{j=1}^{\infty}(\delta_{a_j}+\delta_{b_j})$. The
\underline{topological zeta function} of $\Omega$ with respect to
the sequence $\mathcal{N}$ is
\(\zeta^{\mu_{\Omega}}_{\mathcal{N}}(-\infty,s) \), the multifractal
zeta function of $\mu_{\Omega}$ with respect to $\mathcal{N}$ and
regularity $-\infty$.
\end{dfn}

\ndnt When the open set $\Omega$ has a perfect boundary, there is a
relatively simple breakdown of all the possible multifractal zeta
functions for the measure $\mu_{\Omega}$. Recall that a set is
perfect if it is equal to its set of accumulation points. For
example, the Cantor set is perfect; more generally, all self-similar
sets are perfect (see, e.g., \cite{Falc}). The boundary of a fractal
string is closed; hence, it is perfect if and only if it does not
have any isolated point. The simplicity of the breakdown is due to
the fact that every point-mass is a limit point of other
point-masses. Consequently, the only parameters $\alpha$ that do not
yield identically zero multifractal zeta functions are $\infty$,
$-\infty$ and those which correspond to each length of $\mathcal{N}$
and one or two point-masses.

\begin{thm}\label{thm:perfect}
For a fractal string $\Omega = \bigcup_{j=1}^{\infty}(a_j,b_j)$ with
sequence of lengths $\mathcal{L}$ and perfect boundary, consider
\( \mu_{\Omega}= \sum_{j=1}^{\infty}(\delta_{a_j}+\delta_{b_j}).\)
Suppose that $\mathcal{N}$ is a sequence such that \(l_n>\eta_n \geq
l_{n+1}\) and \(l_n>2\eta_n,\) for all \(n \in \mathbb{N}.\) Then

\[
\zeta^{\mu_{\Omega}}_{\mathcal{N}}(\infty,s) =
\zeta_{\mathcal{L}}(s)
\] and
\[
\zeta^{\mu_{\Omega}}_{\mathcal{N}}(-\infty,s) =
h(s)+\sum_{n=2}^{\infty}m_n(l_n-2\eta_{n})^s,
\]

where $h(s)$ is the entire function given by \(h(s)=
\sum_{p=1}^{k_1(-\infty)}|K^{1}_{p}(-\infty)|^s. \) Moreover, for
every real number $\alpha$ (i.e., for $\alpha \neq \infty,
-\infty$), \( \zeta^{\mu_{\Omega}}_{\mathcal{N}}(\alpha,s) \) is
entire.

\end{thm}

\begin{proof} \( \zeta^{\mu_{\Omega}}_{\mathcal{N}}(\infty,s) =
\zeta_{\mathcal{L}}(s)\) holds by Theorem \ref{thm:recovergzf}.
Since $l_n>2\eta_n$, we have

\[
R^{n}(-\infty)=
\left(\bigcup_{\ell_j>\eta_n}[a_j+\eta_n,b_j-\eta_n]\right)^c.
\]

For $n\geq2$, $J^n(-\infty)$ is made up of $m_n$ intervals of length
$l_n-2\eta_n$ and $2\sum_{p=1}^{n-1}m_{p}$ intervals of length
$\eta_{n-1}-\eta_n$. That is, at each stage $n \geq 2$, we pick up
two $\eta_{n-1}-\eta_n$ terms for each $\ell_j \geq l_{n-1}$ from
the previous stage and one $l_n-2\eta_{n}$ term for each
$\ell_j=l_n$. By construction, the sets $K^n(-\infty)$ do not
include the redundant $\eta_{n-1}-\eta_n$ terms. Therefore,

\[
\zeta^{\mu_{\Omega}}_{\mathcal{N}}(-\infty,s)=
\sum_{p=1}^{k_1(-\infty)}|K^{1}_{p}(-\infty)|^s
+\sum_{n=2}^{\infty}m_n(l_n-2\eta_{n})^s.
\]

\ndnt To prove the last statement in the theorem, note that any
given interval $U \in$ \textbf{X}([0,1]) may contain 0, 1, 2 or
infinitely many endpoints, each of which has a unit point-mass.
Indeed, if $U$ contained an open neighborhood of a point in a
perfect set it would necessarily contain an infinite number of
points. If an interval $U$ contains 1 or 2 endpoints, the
corresponding regularity would appear only at the stage
corresponding to the scale $\eta$ which generated $U$ and perhaps
one more stage. Thus, there are at most two stages contributing
lengths to the multifractal zeta function with the same regularity.
It follows that the multifractal zeta function has finitely many
terms of the form $\ell^s$ where $\ell \in [0,1]$, and hence is
entire.

\end{proof}

\begin{rk} \textnormal{If one were to envision a kind of
multifractal spectrum for the measures $\mu_{\Omega}$ which satisfy
the conditions of Theorem \ref{thm:perfect} in terms of a function
$f(\alpha)$ whose output values are abscissae of convergence of
multifractal zeta functions, the spectrum would be very simple: only
regularity values $\alpha = \pm \infty$ could generate positive
values for $f(\alpha)$. If the weights of the point-masses for some
$\mu_{\Omega}$ were not all the not same, as in the measure $\nu$
from \cite{LVT}, other values of $\alpha$ may be shown to yield
positive $f(\alpha)$ and thus $\nu$ may have a more interesting
multifractal spectrum. See \cite{Rock} for further elaboration on
this perspective.}
\end{rk}

\ndnt For certain fractal strings with perfect boundaries and a
naturally chosen sequence, Theorem \ref{thm:perfect} has the
following corollary.

\begin{cor}\label{cor:perfect}
Assume that $\Omega$ is a fractal string with perfect boundary,
total length 1, and  distinct lengths $\mathcal{L}$ given by
$l_n=ca^{-n}$ with multiplicities $m_n$ for some $a>2$ and $c>0$.
Further, assume that $\mathcal{N}$ is a sequence of scales where
\(\eta_n=l_{n+1}=ca^{-n-1}\), then

\[\zeta^{\mu_{\Omega}}_{\mathcal{N}}(-\infty,s)=f_0(s)
+ f_1(s)\zeta_{\mathcal{L}}(s), \]

where $f_0(s)$ and $f_1(s)$ are entire.
\end{cor}

\begin{proof} By Theorem \ref{thm:perfect},
\begin{eqnarray*}
\zeta^{\mu_{\Omega}}_{\mathcal{N}}(-\infty,s) & = &
    h(s) + \sum_{n=2}^{\infty}m_n(l_n-2l_{n+1})^s\\
    & = &h(s) + c^s\left(\frac{a-2}{a}\right)^s
    \left( \zeta_{\mathcal{L}}(s) - m_1a^{-s} \right).
\end{eqnarray*}
Therefore, the result holds with
\[
f_0(s):= h(s) - m_1c^s \left( \frac{a-2}{a^2}\right)^s \ndnt
\textnormal{and} \ndnt f_1(s):= c^s\left( \frac{a-2}{a} \right)^s.
\]
\end{proof}

\begin{rk} \textnormal{Corollary \ref{cor:perfect} clearly shows
that, in general, the topological zeta functions of the form
\(\zeta^{\mu_{\Omega}}_{\mathcal{N}}(-\infty,s)\) may have poles.
Indeed, since $f_1(s)$ has no zeros, we have that
\(\mathcal{D}^{\mu}_{\mathcal{N}}(-\infty,W) =
\mathcal{D}_{\mathcal{L}}(W) \) for any window $W$.}
\end{rk}

\begin{rk} \textnormal{There are a few key differences between the
result of Theorem \ref{thm:recovergzf} and the results in this
section. For regularity } $\alpha =\infty$, \textnormal{ the form of
the multifractal zeta function is independent of the choice of the
sequence of scales } $\mathcal{N}$ \textnormal{ and the topological
configuration of the fractal string in question. For other
regularity values, however, this is not the case. In particular,
regularity value} $\alpha = -\infty$ \textnormal{ sheds some light
on the topological properties of the fractal string in a way that
depends on } $\mathcal{N}$. \textnormal{This dependence on the
choice of scales is a very common feature in multifractal
analysis.)}
\end{rk}

\ndnt We now define a special sequence that describes the collection of
accumulation points of the boundary of a fractal string $\Omega$.

\s
\begin{dfn} The sequence of \underline{effective lengths} of a
fractal string $\Omega$ with respect to the sequence $\mathcal{N}$
is
\[
\mathcal{K}^{\mu_{\Omega}}_{\mathcal{N}}(-\infty):=\{
|K^n_p(-\infty)| \hs | \hs n \in \mathbb{N}, \hs p \in
\{1,...,k_n(-\infty)\}\},
\]
where $\mu_{\Omega} = \sum_{j=1}^{\infty}(\delta_{a_j}+\delta_{b_j}).$
\end{dfn}

\ndnt This definition is motivated by a key property of the
Hausdorff dimension $dim_H$: it is countably stable, that is,
\[
dim_H(\cup_{n=1}^{\infty} A_n) = \sup_{n \geq 1}dim_H(A_n).
\]
(For this and other properties of $dim_H$, see \cite{Falc}.)
Consequently, countable sets have Hausdorff dimension zero. As such,
countable collections of isolated points do not contribute to the
Hausdorff dimension of a given set. Regularity $-\infty$ picks up
closed intervals of all sizes $\eta_n \in \mathcal{N}$ that contain
an open neighborhood of an accumulation point of the boundary of the
fractal string $\Omega$. The effective sequence (and hence its
multifractal zeta function) describes the gaps between these
accumulation points as detected at all scales $\eta_n \in
\mathcal{N}$, which we now define.

\ndnt The distinct gap lengths are the distinct sums
$g_k:=\sum\ell_{j}$ where $k \in \mathbb{N}$ and the sums are taken
over all $j$'s such that the disjoint subintervals $I_j= (a_j,b_j)$
of $\Omega$ are adjacent and have rightmost and/or leftmost
endpoints (or limits thereof) which are 0, 1 or accumulation points
of $\partial\Omega$. The effective lengths have the following
description: For the scale $\eta_1$, $K^1(-\infty)$ is the union of
the collection of connected components of $R^1(-\infty)$. For
$\eta_n$ such that $n \geq 2$, \( |K^n_p(-\infty)| = g_k - \eta_n\)
if $\eta_n$ is the scale that first detects the gap $g_k$, that is,
if $\eta_n$ is the unique first scale $\eta_k^E$ such that \(
2\eta_{n-1} > g_k \geq 2\eta_n. \)

\ndnt Under appropriate re-indexing, the effective lengths with
multiplicities $m_{E,k}$ (other than $K^1(-\infty)$) are \(
\{l_{E,k}\}_{k \geq 2}\), given by \( l_{E,k}:= g_k - 2\eta_k^E, \)
where the gaps $g_k$ are those such that \(2\eta_1 > g_k\) and the
\(\eta_k^E \in \mathcal{N}_E \subset \mathcal{N}\) are the effective
scales with respect to $\mathcal{N}$ that detect these gaps. The
result is summarized in the next theorem, which gives a formula for
the multifractal zeta function of the measure $\mu_{\Omega}$ with
sequence of scales $\mathcal{N}$ at regularity $-\infty$. The second
formula in Theorem \ref{thm:perfect} above can be viewed as a
corollary to this theorem. Note that the assumption of a perfect
boundary is not needed in the following result.

\begin{thm}\label{thm:el}
For a fractal string $\Omega$ with sequence of lengths $\mathcal{L}$
and for a sequence of scales $\mathcal{N}$ such that \( \eta_n
\searrow 0\), the topological zeta function is given by

\begin{eqnarray*}
\zeta^{\mu_{\Omega}}_{\mathcal{N}}(-\infty,s) & = &
\sum_{p=1}^{k_1(-\infty)}|K^{1}_{p}(-\infty)|^s
+\sum_{k=1}^{\infty}m_{E,k}l_{E,k}^s,
\end{eqnarray*}

for \textnormal{Re}$(s)$ large enough.
\end{thm}

\ndnt The next section investigates the application of the results
of Sections \ref{rvigzf} and \ref{rvnitzf} to the Cantor String, as
defined in Section \ref{fsgzf}, and the variants thereof.

\section{Variants of the Cantor String}\label{cs}

Let $\mathcal{L}$ be the sequence of lengths in the complement of
the Cantor set, which is also known as the Cantor String $\Omega_1$.
(See Example \ref{eg:cs} and Figure 2.) Then $l_n=3^{-n}$ and
$m_n=2^{n-1}$ for all $n$. We will discuss three examples of fractal
strings  involving this sequence of lengths, but for now consider
the following one.

\ndnt Let $\Omega_2$ be the open subset of $[0,1]$ whose lengths are
also $\mathcal{L}$ but arranged in non-increasing order from right
to left, as in Example \ref{eg:cs2}. That is, the only accumulation
point of $\partial\Omega_2$ is $0$ (see Figures 4, 6 and 7). In each
figure, portions of the approximation of the string that appear
adjacent are actually separated by a single point in the support of
the measure. Gaps between the different portions and the points 0
and 1 contain the smaller portions of the string, isolated endpoints
and accumulation points of endpoints.

\begin{figure}
    \epsfysize=9cm\epsfbox{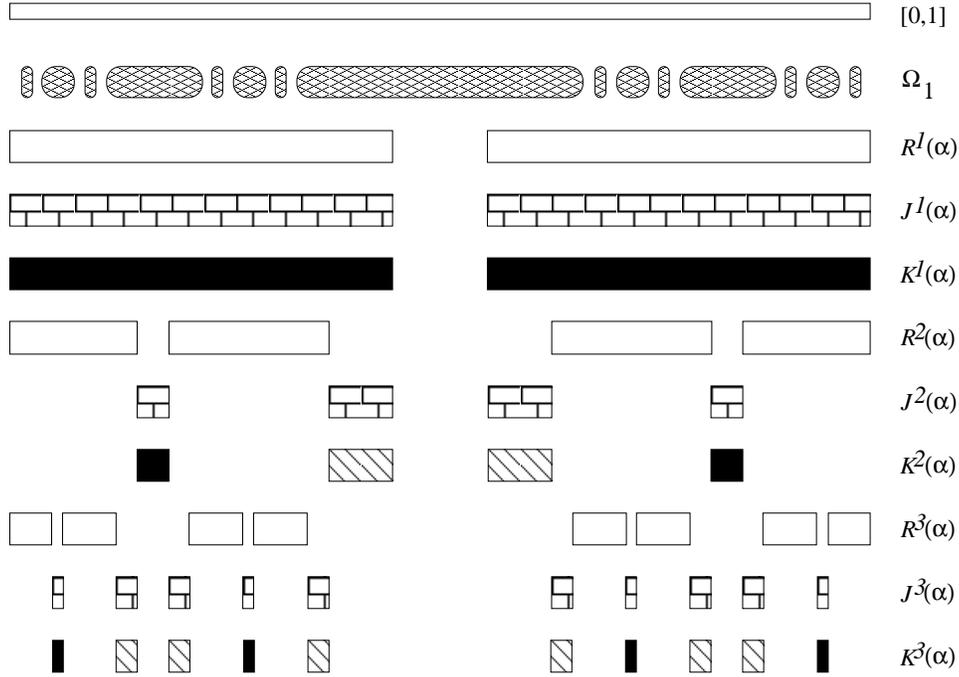}
    \caption{\textit{The first three stages in the construction of
the topological zeta function of }$\Omega_1$,
$\zeta^{\mu_1}_{\mathcal{N}}(-\infty,s),$ \textit{where}
$\mathcal{N}$ \textit{is the set of distinct lengths of the Cantor
String beginning with} 1/9.}
\end{figure}

\ndnt Consider the following measures which have singularities on a
portion of the boundary of $\Omega_1$ and $\Omega_2$, respectively:
$\mu_q=\mu_{\Omega_q}$, with $q = 1$ or $2$, where $\mu_{\Omega_q}$
is defined as in Section \ref{rvnitzf}. These measures have a unit
point-mass at every endpoint of the intervals which comprise
$\Omega_1$ and $\Omega_2$, respectively.

\begin{figure}
    \epsfysize=9cm\epsfbox{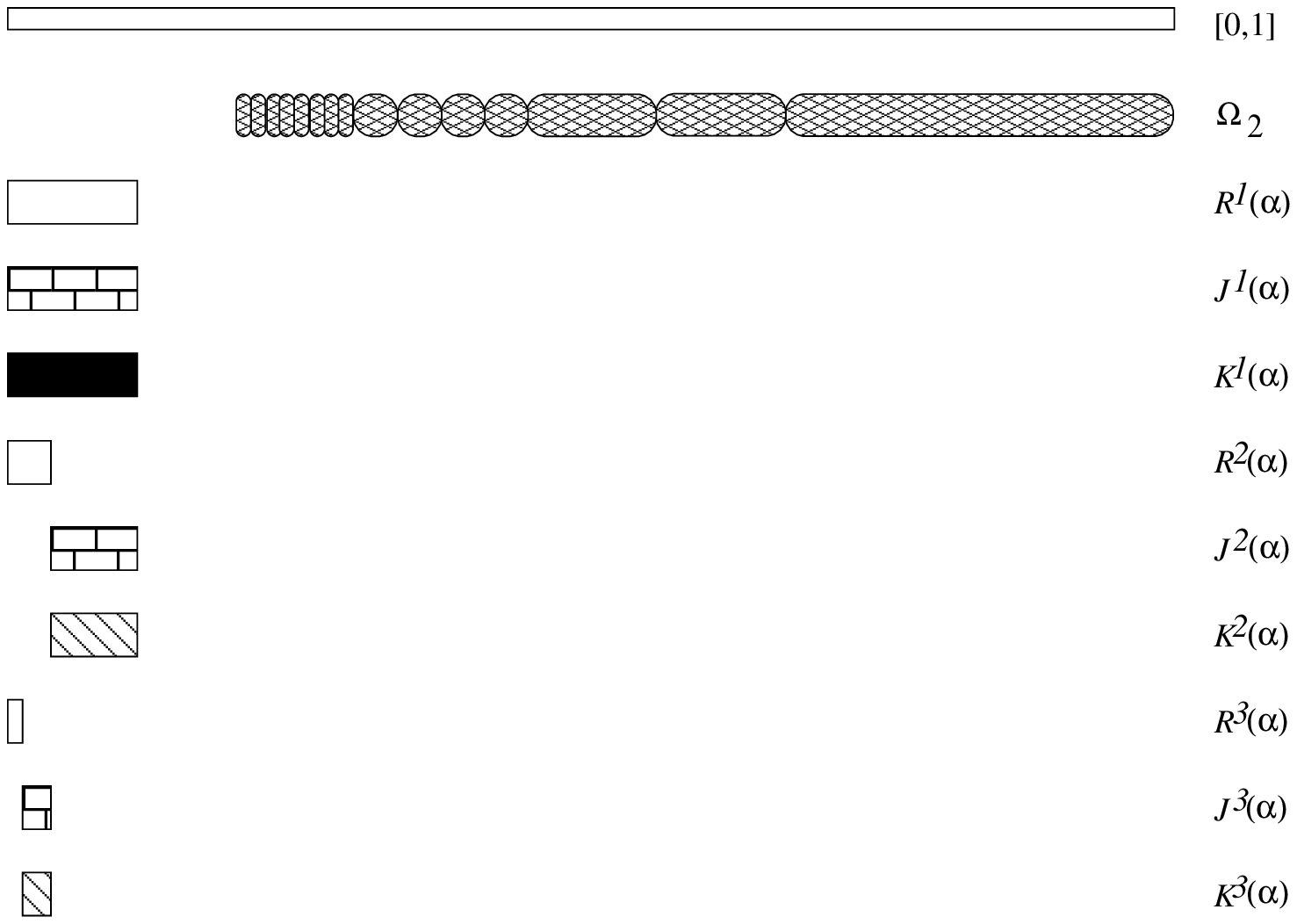}
    \caption{\textit{The first three stages in the construction of
the topological zeta function of }$\Omega_2$,
$\zeta^{\mu_2}_{\mathcal{N}}(-\infty,s),$ \textit{where}
$\mathcal{N}$ \textit{is the set of distinct lengths of the Cantor
String beginning with} 1/9.}
\end{figure}

\ndnt Let $\mathcal{N}$ be such that \(l_n>\eta_n \geq l_{n+1}\) and
\(l_n>2\eta_n\). Such sequences exist for the Cantor String. For
instance, $\forall n \in \mathbb{N}$, let
\(\eta_n=l_{n+1}=3^{-n-1}.\) Theorem \ref{thm:recovergzf} yields
\[
\zeta^{\mu_1}_{\mathcal{N}}(\infty,s)=
\zeta^{\mu_2}_{\mathcal{N}}(\infty,s)= \zeta_{CS}(s).
\]

\ndnt When $\alpha = -\infty$ the topological zeta functions for
$\Omega_1$ and $\Omega_2$ are, respectively,
\[
\zeta^{\mu_1}_{\mathcal{N}}(-\infty,s) = 2(l_{1}+\eta_1)^s
    +\sum_{n=2}^{\infty}2^{n-1}(l_n-2\eta_{n})^s
\]
and
\[
\zeta^{\mu_2}_{\mathcal{N}}(-\infty,s) = \eta_{1}^s.
\]

In either case,
\begin{eqnarray*}
    -\infty = A(U)
    & = & \frac{\sum_{a_j,b_j \in U}1}{\log|U|}
\end{eqnarray*}
if and only if
\[
\# \{j \hs | \hs a_j \in U\}+\#\{j \hs | \hs b_j \in U\}=\infty.
\]

\ndnt In the case of $\mu_2$, the only closed interval of length
$\eta_n$ that contains infinitely many unit point-masses is
$[0,\eta_n]$. So,
\[
R^n(-\infty)=[0,\eta_n]
\]
which means
\[
J^1(-\infty)= K^1(-\infty)=[0,\eta_1]
\] and for $n\geq 2$,
\[
J^n(-\infty)=(\eta_n,\eta_{n-1}].
\]

All of the terms from  $J^n(-\infty)=(\eta_n,\eta_{n-1}]$ are
redundant. Therefore,
\[
K^n(-\infty) = \emptyset
\]
and
\[
\zeta^{\mu_2}_{\mathcal{N}}(-\infty,s) = \eta_{1}^s.
\]

\ndnt The case of $\mu_1$ for regularity $\alpha=-\infty$ is more
complicated and is a result of Theorem \ref{thm:perfect}. This is
due to the fact that every point-mass is a limit point of other
point-masses. That is, the Cantor set is a perfect set, thus
Corollary \ref{cor:perfect} applies when $\mathcal{N}$ is chosen so
that $\eta_n = 3^{-n-1}$ for all $n \in \mathbb{N}$.

\begin{rk}\label{rk:cs2}
\textnormal{Clearly, for every $\mathcal{N}$ chosen as above in the
discussion of $\mu_2$, $\mathcal{D}^{\mu_2}_{\mathcal{N}}(-\infty)$
is empty. In contrast, it follows from the above discussion that it
is easy to find a sequence $\mathcal{N}$ such that
$\mathcal{D}^{\mu_1}_{\mathcal{N}}(-\infty)$ is non-empty and even
countably infinite.}
\end{rk}

\ndnt Shortly we will consider another fractal string, $\Omega_3$,
in addition to the Cantor String $\Omega_1$ and the string
$\Omega_2$. All of these fractal strings have the same sequence of
lengths. As such, these strings all have the same Minkowski
dimension, namely $\log_{3}{2}$. However, their respective Hausdorff
dimensions do not coincide, a fact that is detected by the
topological zeta functions but the theory of fractal strings
developed in \cite{LapvF1,LapvF4} does not describe. For a certain,
natural choice of sequence of scales $\mathcal{N}$, the topological
zeta functions of the fractal strings $\Omega_q$ (as above) have
poles on a discrete line above and below the Hausdorff dimension of
the boundaries of these fractal strings (see Figures 4 and 6--8). In
\cite{LapvF1,LapvF4} it is shown that the \textit{complex
dimensions} of the fractal strings $\Omega_q$, for $q=1,2,3$ are

\[
\mathcal{D}_{CS} = \left\{\log_3{2}+\frac{2i\pi m}{\log3} \hs | \hs
m \in \mathbb{Z}\right\}.
\]

These are the poles of
\[
\zeta^{\mu_q}_{\mathcal{N}}(\infty,s)=
\zeta_{CS}(s)=\frac{3^{-s}}{1-2\cdot3^{-s}}.
\]
(See Section \ref{fsgzf} above.) As noted earlier, the geometric
zeta function of the Cantor String does not see any difference
between the open sets $\Omega_q$, for $q=1,2,3$. However, the
multifractal zeta functions of the measures $\mu_q$ with the same
such $\mathcal{N}$ and regularity $\alpha =-\infty$ are quite
different. For the remainder of this section, unless explicitly
stated otherwise, we choose $\mathcal{N} =
\{3^{-n-1}\}_{n=1}^{\infty}$.

\ndnt We now consider more specifically the fractal string
$\Omega_3$ mentioned above. This fractal string is comprised of a
Cantor-like string and an isolated accumulation point at 1. The
lengths comprising the Cantor-like string are constructed by
connecting two intervals with consecutive lengths. The remaining
lengths are arranged in non-increasing order from left to right,
accumulating at 1. That is, for $n \geq 1$, the gap lengths are
$3^{-2n+1}+3^{-2n} = 4\cdot3^{-2n}$ with multiplicities $2^{n-1}$
and therefore the effective lengths are $2\cdot3^{-2n}$ with
multiplicities $2^{n-1}$. (See Figure 8.)

\begin{figure}
    \epsfysize=9cm\epsfbox{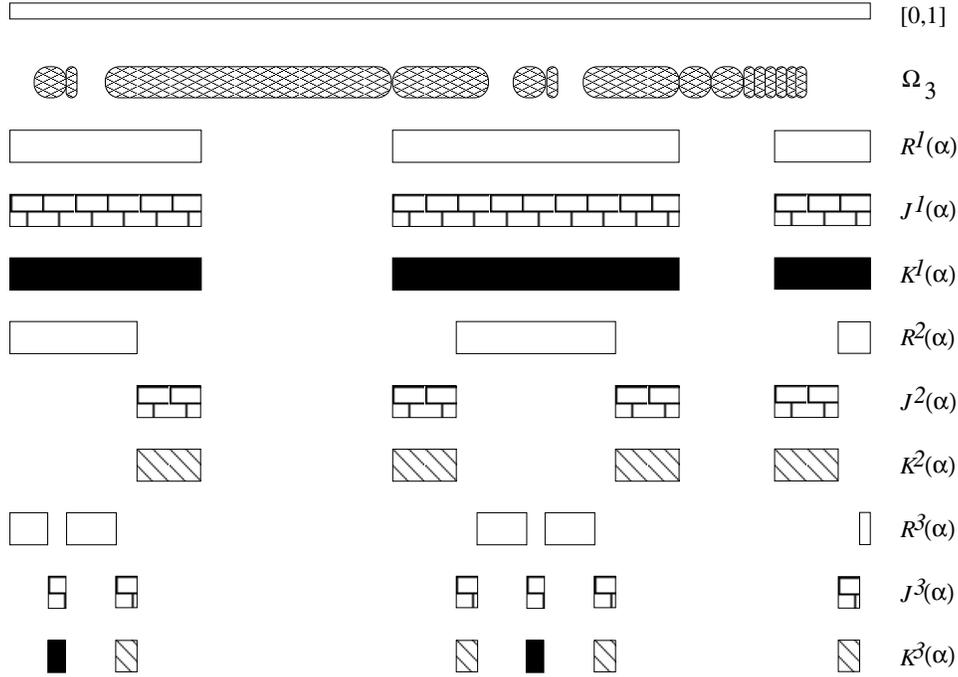}
    \caption{\textit{The first three stages in the construction of
the topological zeta function of }$\Omega_3$,
$\zeta^{\mu_3}_{\mathcal{N}}(-\infty,s),$ \textit{where}
$\mathcal{N}$ \textit{is the set of distinct lengths of the Cantor
String beginning with} 1/9.}
\end{figure}

\ndnt The Hausdorff dimension of the boundary of each fractal string
$\Omega_q$ ($q=1,2,3$) is easily determined. For a set $F$, denote
the Hausdorff dimension by $dim_H(F)$ and the Minkowski dimension by
$dim_M(F)$. We have, for $q=1,2,3$:

\begin{eqnarray*}
dim_H(\partial\Omega_1) = dim_M(\partial\Omega_q) &=& \log_{3}{2},\\
dim_H(\partial\Omega_2) &=& 0,\\
dim_H(\partial\Omega_3) &=& \log_{9}{2}.
\end{eqnarray*}

The first line in the displayed equation above holds since the
Minkowski dimension depends only on the lengths of the fractal
strings and, furthermore, the Cantor set $\partial\Omega_1$ is a
strictly self-similar set whose similarity transformations satisfy
the \textit{open set condition}, as defined, for example, in
\cite{Falc}. Thus, the Minkowski and Hausdorff dimensions coincide
for $\partial\Omega_1$. The second equality holds because
$\partial\Omega_2$ is a countable set. The third holds because
$\partial\Omega_3$ is the disjoint union of a strictly self-similar
set and a countable set, and Hausdorff dimension is (countably)
stable. We justify further below.

\ndnt Theorem \ref{thm:perfect}, Corollary \ref{cor:perfect}, and
Theorem \ref{thm:el} will be used to generate the following closed
forms of the zeta functions
$\zeta^{\mu_q}_{\mathcal{N}}(-\infty,s).$

\ndnt For the Cantor String $\Omega_1$ and the corresponding measure
$\mu_1$, we have by Corollary \ref{cor:perfect},

\begin{eqnarray*}
\zeta^{\mu_1}_{\mathcal{N}}(-\infty,s) &=&
    2\left( \frac{1}{3}+\frac{1}{9} \right)^s
    +\sum_{n=2}^{\infty} 2^{n-1} \left(
    \frac{1}{3^n}-\frac{2}{3^{n+1}} \right)^s\\
&=& 2\left( \frac{4}{9} \right)^s
    +\frac{2}{27^s} \left(\frac{1}{1-2\cdot3^{-s}} \right).
\end{eqnarray*}

The poles of $\zeta^{\mu_1}_{\mathcal{N}}(-\infty,s)$ are the same
as the poles of the geometric zeta function of the Cantor String.
They are given by

\[
\mathcal{D}^{\mu_1}_{\mathcal{N}}(-\infty) = \left\{ \log_3{2} +
\frac{2i\pi m}{\log3} \hs | \hs m \in \mathbb{Z} \right\} =
\mathcal{D}_{CS}.
\]

\begin{rk}\label{rk:merocont} \textnormal{Note that the above computation
of }$\zeta^{\mu_1}_{\mathcal{N}}(-\infty,s)$ \textnormal{is
justified, }a priori\textnormal{, for Re}$(s) >
\log_{3}2.$\textnormal{ However, by analytic continuation, it
clearly follows that
}$\zeta^{\mu_1}_{\mathcal{N}}(-\infty,s)$\textnormal{ has a
meromorphic continuation to all of }$\mathbb{C}$\textnormal{ and is
given by the same resulting expression for every }$s \in
\mathbb{C}.$\textnormal{ Analogous comments apply to similar
computations elsewhere in the paper.}
\end{rk}

\ndnt Since $\partial\Omega_2$ has only one accumulation point,
there is only one term in the corresponding topological zeta
function for $\Omega_2$. We immediately have

\[
\zeta^{\mu_2}_{\mathcal{N}}(-\infty,s) = \frac{1}{9^s},
\]

which, of course, is entire and has no poles.

\ndnt For $\Omega_3$, we have

\begin{eqnarray*}
\zeta^{\mu_3}_{\mathcal{N}}(-\infty,s) &=& h_3(s)
    +\sum_{n=2}^{\infty}m_n \left(l_{2n-1}+l_{2n}
    -2\eta_{2n-1}\right)^s\\
&=& h_3(s)+ \left(\frac{2^{s+1}}{81^s}\right) \left(
    \frac{1}{1-2\cdot9^{-s}} \right),\\
\end{eqnarray*}

where $h_3(s)$ is entire. Therefore, the poles of
$\zeta^{\mu_3}_{\mathcal{N}}(-\infty,s)$ are given by
\[
\mathcal{D}^{\mu_3}_{\mathcal{N}}(-\infty) = \left\{ \log_9{2}+
\frac{2i\pi m}{\log9} \hs | \hs m \in \mathbb{Z} \right\}.
\]

\ndnt Let us summarize the results of this section. We chose the
sequence of scales $\mathcal{N}$ to be
$\{3^{-n-1}\}_{n=1}^{\infty}$. For $q=1,2,3$, the multifractal zeta
function of each measure $\mu_q$ with regularity $\alpha =\infty$ is
equal to the geometric zeta function of the Cantor String, as
follows from Theorem \ref{thm:recovergzf}. Thus, obviously, the
collections of poles $\mathcal{D}^{\mu_q}_{\mathcal{N}}(\infty)$
each coincide with the complex dimensions of the Cantor String.

\ndnt For regularity $\alpha = -\infty$, the multifractal zeta
functions are the topological zeta functions for the fractal strings
$\Omega_q$. The respective sets of complex dimensions
$\mathcal{D}_{\mathcal{N}}^{\mu_q}(-\infty)$ differ for each
$q=1,2,3$. Specifically,
$\mathcal{D}_{\mathcal{N}}^{\mu_1}(-\infty)$ is exactly the same set
of poles as $\mathcal{D}_{\mathcal{N}}^{\mu_1}(\infty)$ (and
$\mathcal{D}_{CS}$), corresponding to the fact that
$\partial\Omega_1$ has equal Minkowski and Hausdorff dimensions.
Furthermore, since $\zeta^{\mu_2}_{\mathcal{N}}(-\infty,s)$ is an
entire function, $\mathcal{D}_{\mathcal{N}}^{\mu_2}(-\infty)$ is the
empty set. Additionally, $\partial\Omega_2$ has Hausdorff dimension
equal to zero. Finally, $\mathcal{D}_{\mathcal{N}}^{\mu_3}(-\infty)$
is a discrete line of poles above and below the Hausdorff dimension
of $\partial\Omega_3$, which is $\log_9{2}$. In all of these cases,
the multifractal zeta functions with regularity $-\infty$ and their
corresponding poles depend heavily on the choice of sequence of
scales $\mathcal{N}$.

\ndnt This section further illustrates the dependence of the
multifractal zeta function with regularity $\alpha = -\infty$ on the
topological configuration of the fractal string in question as well
as the choice of scales $\mathcal{N}$ used to examine the fractal
string. As before, following Theorem \ref{thm:recovergzf},
regularity $\alpha =\infty$ corresponds to a multifractal zeta
function that depends only on the lengths of the fractal string in
question.

\section{Concluding Comments}\label{conclusion}

The main object defined in this paper, the multifractal zeta
function, was originally designed to provide a new approach to
multifractal analysis of measures which exhibit fractal structure in
a variety of ways. In the search for examples with which to work,
the authors found that the multifractal zeta functions can be used
to describe some aspects of fractal strings that extend the existing
notions garnered from the theory of geometric zeta functions and
complex dimensions of fractal strings developed in
\cite{LapvF1,LapvF4}.

\ndnt Regularity value $\alpha = \infty$ has been shown to precisely
recover the geometric zeta function of the complement in [0,1] of
the support of a measure which is singular with respect to the
Lebesgue measure. This recovery is independent of the topological
configuration of the fractal string that is the complement of the
support and occurs under the mild condition that the sequence of
scales $\mathcal{N}$ decreases to zero. The fact that the recovery
does not depend on the choice of sequence $\mathcal{N}$ (as long as
it decreases to zero) is unusual in multifractal analysis.

\ndnt Regularity value $\alpha = -\infty$ has been shown to reveal
more topological information about a given fractal string by using a
specific type of measure whose support lies on the boundary of the
fractal string. The results depend on the choice of sequence of
scales $\mathcal{N}$ (as is generally the case in multifractal
analysis) and the topological structure inherent to the fractal
string. Moreover, the topological configuration of the fractal
string is illuminated in a way which goes unnoticed in the existing
theories of fractal strings, geometric zeta functions and complex
dimensions, such as the connection to the Hausdorff dimension.

\ndnt We next point out several directions for future research, some
of which will be investigated in later papers:

\ndnt Currently, examination of the families of multifractal zeta
functions for truly multifractal measures on the real line is in
progress, measures such as the binomial measure and mass
distributions which are supported on the boundaries of fractal
strings. Preliminary investigation of several examples suggests that
the present definition of the multifractal zeta functions may need
to be modified in order to handle such measures. Such changes take
place in \cite{LR,LR2,LVM,Rock} and are discussed briefly in the
next section.

\ndnt In the longer term, it would also be interesting to
significantly modify our present definitions of multifractal zeta
functions in order to undertake a study of higher-dimensional
fractals and multifractals. A useful guide in this endeavor should
be provided by the recent work of Lapidus and Pearse on the complex
dimensions of the Koch snowflake curve (see \cite{LapPe1}, as
summarized in \cite{LapvF4}, \S 12.3.1) and more generally but from
a different point of view, on the zeta functions and complex
dimensions of self-similar fractals and tilings in $\mathbb{R}^d$
(see \cite{LapPe2,LapPe3} and \cite{Pearse}, as briefly described in
\cite{LapvF4}, \S 12.3.2, along with the associated tube formulas).

\ndnt In \cite{HL}, the beginning of a theory of complex dimensions
and random zeta functions was developed in the setting of random
fractal strings. It would be worth extending the present work to
study random multifractal zeta functions, first in the same setting
as \cite{HL}, and later on, in the broader framework of random
fractals and multifractals considered, for example, in
\cite{AP,Falc,GrMouWi,Ja3,LVM,Man,Ol}.

\ndnt These are difficult problems, both conceptually and
technically, and they will doubtless require several different
approaches before being successfully tackled. We hope, nevertheless,
that the concepts introduced and results obtained in the present
paper can be helpful to explore these and related directions of
research.

\section{Epilogue}\label{epilogue}

This epilogue focuses on some of the recent results presented in
\cite{LR,LR2,LVM,Rock}, much of which was motivated by this paper
and established concurrently or subsequently. In particular, we
discuss results on the multifractal analysis of the binomial measure
$\beta$ mentioned in Section \ref{ma}, specifically the
reformulation of the classical multifractal spectrum $f(\alpha)$ as
described in \cite{Falc}, for instance.

\begin{figure}
    \epsfysize=4cm\epsfbox{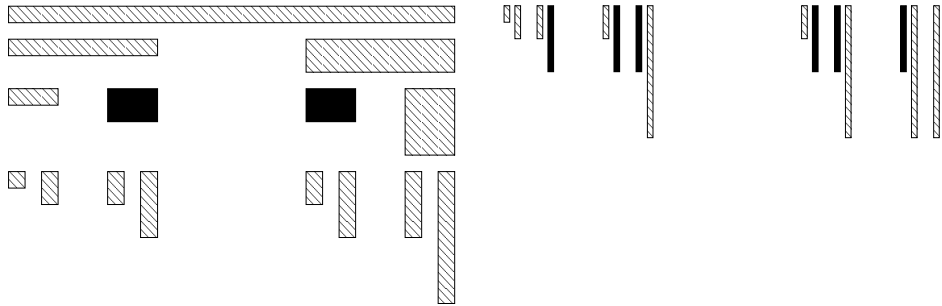}
    \caption{\textit{The first stages in the construction of
the partition zeta function
$\zeta^{\beta}_{\mathfrak{P}}(\alpha(1,2),s)$ whose regularity
$\alpha = \alpha(1,2)$ yields the maximum value of the multifractal
spectrum $f(\alpha)$ of the binomial measure $\beta$. The solid
black bars contribute their lengths to the series formula of the
partition zeta function.}}
\end{figure}

\ndnt In \cite{LR,LR2,Rock}, a new family of zeta functions
parameterized by a countable collection of regularity values called
{\it partition zeta functions} are defined and considered. These
functions were inspired by both the multifractal and the geometric
zeta functions. Partition zeta functions are easier to define than
multifractal zeta functions and produce the desired result of
reformulating the multifractal spectrum $f(\alpha)$ for the binomial
measure $\beta$, for instance. (See Example \ref{eg:mmcs} along with
Figure 9 for a description of $\beta$ and its multifractal
spectrum.) The partition zeta functions of $\beta$ with respect to
the family of weighted partitions $\mathfrak{P}$ used to help define
$\beta$ are of the following form:
\[
\zeta^{\beta}_{\mathfrak{P}}(\alpha(k_1,k_2),s) =
\sum_{n=1}^{\infty} \binom{nk_2}{nk_1} 3^{-k_2ns},
\]
where the regularity $\alpha = \alpha(k_1,k_2)$ is parameterized by
scale and weight in terms of nonnegative integers $k_1$  and $k_2$,
$s$ is in $\mathbb{C}$ accordingly, and $\binom{nk_2}{nk_1}$ are
binomial coefficients. See Figure 9 for the construction in the case
of regularity $\alpha = \alpha(1,2)$ which yields the maximum value
of the spectrum (which is also the Minkowski dimension of the
support of $\beta$). The value of the function $f(\alpha)$, in this
case, is defined as the abscissa of convergence of
$\zeta^{\beta}_{\mathfrak{P}}(\alpha,s)$. Various generalizations of
the present example can be treated in a similar manner.

\ndnt In \cite{LVM}, the {\it modified multifractal zeta function}
(among other zeta functions) is defined and its properties are
investigated. As with the partition zeta function, the definition
depends on a given measure $\mu$ and a sequence of partitions
$\mathcal{P}_n$, which corresponds to a natural family of partitions
in the case of the binomial measure $\beta$ (a similar comment
applies to the more general multinomial measures). This zeta
function has the following form:
\[
\zeta(q,s) = \sum_{n=1}^{\infty} \sum_{U \in \mathcal{P}_n} \mu(U)^q
|U|^s.
\]
When $\mu$ is a self-similar probability measure with weights $p_j$
and scales $r_j$ (for $j=1,...,J$), the modified multifractal zeta
function becomes:
\[
\zeta(q,s) = \frac{1}{1-\sum_{j=1}^J p_j^q r_j^s}.
\]
For fixed $q \in \mathbb{R}$, the negative of the abscissa of
convergence $\sigma(q)$ of $\zeta(q,s)$ is the Legendre transform of
the multifractal spectrum of $\mu$. Additionally, in the spirit of
the theory of complex dimensions of \cite{LapvF1,LapvF4}, the poles
of the modified multifractal zeta function allow for the detection
and measurement of the possible oscillatory behavior of the
so-called continuous partition function in the self-similar case.

\ndnt Overall, the use of a zeta function or families of zeta
functions to extrapolate information regarding multifractal measures
appears to be quite a promising prospect. The study of the geometric
and topological zeta functions of fractal strings has shown what may
lay ahead for similar investigations in multifractal analysis.

\as

\ndnt \textsc{Michel L. Lapidus},\\
{\tiny \textsc{Department of Mathematics, University of California,
Riverside, CA} 92521-0135, \textsc{USA} \par}

\ndnt \textit{E-mail address:} \textbf{lapidus@math.ucr.edu}

\s

\ndnt \textsc{Jacques L\'{e}vy V\'{e}hel},\\
{\tiny \textsc{Projet Fractales, INRIA Rocquencourt, B. P. 105, Le
Chesnay Cedex, France}
\par}

\ndnt \textit{E-mail address:} \textbf{Jacques.Levy\_Vehel@inria.fr}

\s

\ndnt \textsc{John A. Rock},\\
{\tiny \textsc{Department of Mathematics, California State
University, Stanislaus, Turlock, CA} 95382, \textsc{USA} \par}

\ndnt \textit{E-mail address:} \textbf{jrock@csustan.edu}

\end{document}